\documentstyle[array,epsfig,preprint,prl,aps]{revtex}

\begin{document}
\draft


\title{Entanglement induced by a single-mode heat environment}
\author{M. S. Kim,$^1$ Jinhyoung Lee,$^2$ D. Ahn$^2$ and P. L. Knight$^3$} \address{$^1$
  School of Mathematics and Physics, The Queen's
  University, Belfast \\ BT7 1NN, United Kingdom\\
  $^2$ Creative Research Institute of Quantum Information Processing
  and Systems, \\ The University of Seoul, Seoul,
  Korea \\
  $^3$ Optics Section, The Blackett Laboratory, Imperial College, London SW7 2BW, United Kingdom}

\date{\today}

\maketitle

\begin{abstract}
  A thermal field, which frequently appears in problems of
  decoherence, provides us with minimal information about the field.
  We study the interaction of the thermal field and a quantum system
  composed of two qubits and find that such a chaotic field with 
  minimal information can nevertheless entangle the qubits which are prepared initially in a
  separable state.  This simple model of a quantum register
  interacting with a noisy environment allows us to understand how memory of the
  environment affects the state of a quantum register.
\end{abstract}

\pacs{PACS number(s); 03.65.Bz, 03.67, 42.50.Dv}


\newpage

A thermal field is emitted by a source in thermal equilibrium at
temperature $T$.  The thermal field is the field about which we have 
minimal information, as we know only the mean value of the energy
\cite{Barnett97}.  It arises frequently in problems involving the
coupling of a system to its environment which is in thermodynamic
equilibrium.  Recently, Bose {\it et al.} \cite{Bose01} showed that entanglement can
always arise in the interaction of a single qubit in a pure state with
an arbitrarily large system in any mixed state and illustrated this using a model of
the interaction
of a two-level atom with a thermal field.  Using this model, they
studied the possibility of entangling a qubit with a large system
defined in an infinite dimensional Hilbert space.  The entanglement
between the system and the thermal field reduces the system to a mixed state when the
field variables are traced over.  In this paper we address the question
`Is it possible for a thermal field, which is a highly chaotic
field, to induce entanglement between qubits?'  The entanglement of an
atom and a field may be a totally different problem from the
entanglement of two atoms (or qubits) by their mutual interaction with
a chaotic field.

We consider a quantum register composed of two two-level atoms interacting
with a single-mode thermal field.  We will investigate the
entanglement between the two atoms, which are initially separable.
This simple interaction model of a quantum register with its environment
allows us to understand how the memory of the environment affects the
state of a quantum register.

When a quantum system of two qubits prepared in $\hat{\rho}_s$
interacts with an environment represented by the density operator
$\hat{\rho}_E$, the system and environment evolve
for a finite time, governed by the unitary time evolution operator
$\hat{U}(t)$.  The density operator for the system and environment at
time $t$ is
\begin{equation}
\hat{\rho}(t)= \hat{U}(t)(\hat{\rho}_E\otimes\hat{\rho}_s)\hat{U}^\dag(t).
\label{total-system-t}
\end{equation}
After performing a partial trace over environment variables, we find
the final density matrix $\hat{\rho}_s(t)$ of the quantum system in
the following Kraus representation \cite{Nielsen}
\begin{equation}
\label{Kraus-representation}
\hat{\rho}_s(t)=\mbox{Tr}_E\hat{\rho}(t)=\sum_\mu \hat{K}_\mu\hat{\rho}_s\hat{K}_\mu^\dag
\end{equation}
where the Kraus operators $\hat{K}_\mu$'s satisfy the property:
$\sum_\mu \hat{K}_\mu^\dag \hat{K}_\mu= \openone_s$.  Unitary
evolution of the quantum system is a special case in which there is
only one non-zero term in the operator sum
(\ref{Kraus-representation}).  If there are two or more terms, the
pure initial state becomes mixed.  For the mixed thermal environment,
not only entanglement but also classical correlation of the system and
environment can make the system evolve from the initial pure state into a mixed one.
The mutual
information between the system and the environment is non-zero if the information for the total system is not
the same as the total sum of information for each subsystem.  A
non-zero mutual information is due to classical correlation and/or
entanglement \cite{LeeKim00}.  Thus there being non-zero mutual
information between the system and environment is a sufficient
condition for the initial pure system to have evolved into a mixed
state.
The action of $\hat{K}_\mu$ projects the system into a pure state of
$\hat{K}_\mu\hat{\rho}\hat{K}_\mu^\dag$ when the initial state is pure. 

The Jaynes-Cumming model, describing the essential physics of the
interaction of a single-mode radiation field with a two-level atom,
has provided an excellent test ground for quantum mechanics
\cite{Jaynes-Cummings,ShoreKnight93}.  In this paper, we are
interested in the interaction of two two-level atoms with a single-mode
thermal field, whose annihilation and creation operators are denoted by
$\hat{a}$ and $\hat{a}^\dag$.  In the context of cavity quantum
electrodynamics, the study of the interaction with atoms has been
used to study for the generation of entanglement between two atoms when the
two atoms are present simultaneously in the cavity \cite{Plenio99} and when the two
atoms interact consecutively with the cavity \cite{Kud}.  In these
studies, the cavity is normally prepared in the vacuum and by the
superposition of processes involving depositing and not depositing one photon into the
cavity, the two atoms can evolve into their entangled
state.  However we are interested in the possibility of entanglement via
a chaotic thermal field about which we have only minimum
information.

For simplicity, we consider the identical two-level atoms 1 and 2 which are coupled to a single-mode
thermal field with the same coupling constant $\gamma$.  The ground
and excited states for the atom $i$ ($i=1,2$) are, respectively,
denoted by $|g\rangle_i$ and $|e\rangle_i$.  The cavity mode is
assumed to be resonant with the atomic transition frequency.  Under
the rotating wave approximation, the Hamiltonian in the interaction
picture is
\begin{equation}
\label{Hamiltonian}
\hat{H}_I=\hbar \gamma\sum_{i=1,2}\left(\hat{a}\hat{\sigma}_i^+
  +\hat{a}^\dag\hat{\sigma}_i^- \right)
\end{equation}
where the atomic transition operators are 
$\hat{\sigma}_i^-=|g\rangle_i\langle e|$ and
$\hat{\sigma}_i^+=|e\rangle_i\langle g|$.

The density operator for the combined atom-field system follows a
unitary time evolution generated by the evolution operator,
$\hat{U}(t)=\exp(-i\hat{H}t/\hbar)$.  The evolution operator has been
extensively studied for a two- and three-level atom Jaynes-Cummings
model \cite{ShoreKnight93}.  For the interaction of two two-level
atoms with a single-mode field, using Hamiltonian (\ref{Hamiltonian})
and Taylor expansions of sine and cosine functions, we find that the
analytical form of the evolution operator is given in the atomic basis
$\{|ee\rangle,|eg\rangle,|ge\rangle,|gg\rangle\}$ by
\begin{eqnarray}
\label{unitary-operator}
  \hat{U}(t) = \left(
    \begin{array}{cccc}
      2\gamma^2 \hat{a} (\hat{C}-\hat{\Theta})\hat{a}^\dagger+1 & -i
      \gamma \hat{a} \hat{S} &  -i \gamma \hat{a} \hat{S} & 
      2\gamma^2 \hat{a}(\hat{C}-\hat{\Theta}) \hat{a}\\
      -i\gamma \hat{S} \hat{a}^\dagger & {1\over 2}(\cos \hat{\Omega}
      t +1)& {1\over 2}(\cos \hat{\Omega} t -1) &  
        -i\gamma \hat{S} \hat{a}\\
      -i\gamma \hat{S} \hat{a}^\dagger & {1\over 2}(\cos \hat{\Omega}
      t -1)& {1\over 2}(\cos \hat{\Omega} t +1) &  
        -i\gamma \hat{S} \hat{a}\\
      2\gamma^2\hat{a}^\dagger(\hat{C}-\hat{\Theta})\hat{a}^\dagger &
      -i \gamma 
      \hat{a}^\dagger \hat{S}& -i \gamma\hat{a}^\dagger \hat{S}
      &2\gamma^2\hat{a}^\dagger(\hat{C}-\hat{\Theta})\hat{a}+1   
    \end{array}
    \right)
\end{eqnarray}
where $\hat{\Omega}^2=\hat{\Theta}^{-1}=2\gamma^2
(2\hat{a}^\dag\hat{a}+1)$ and the time-dependent operators $\hat{C}$
and $\hat{S}$ are defined by
\begin{equation}
\hat{C}=\hat{\Theta} \cos\hat{\Omega} t \mbox{~~and~~}
\hat{S}=\hat{\Omega}^{-1}\sin\hat{\Omega} t. 
\end{equation}

The thermal radiation field with its mean photon number $\bar n$ is a
weighted mixture of Fock states and its density operator is
represented by $\hat{\rho}_E=\sum_n P_n|n\rangle\langle n|$ where the
weight function $P_n$ is
\begin{equation}
\label{thermal-density}
P_n={\bar n^n\over(1+\bar n)^{n+1}}.
\end{equation}
We are interested in the evolution of the quantum system hence the
time-dependent density operator $\hat{\rho}_s(t)$ is obtained by
tracing over the field variables as in
Eq.(\ref{Kraus-representation}): $ \hat{\rho}_s(t)=\sum_{nm}P_n\langle
m|\hat{U}(t)|n\rangle\hat{\rho}_s(0)\langle
n|\hat{U}^\dag(t)|m\rangle.  $ We denote the $ij$th element of the
matrix $\langle m|(\hat{U})|n\rangle$ by $U_{ij}^{nm}$.  Using the
orthogonality of the Fock state, $\hat{\rho}_s(t)$ is obtained
in the Kraus representation:
\begin{equation}
\label{Kraus-representation-1}
\hat{\rho}_s(t)=\sum_{n=0}^\infty\sum_{\mu=1}^{5}P(n)\hat{K}_\mu^n\hat{\rho}_s(0)
\hat{K}_\mu^{n\dag}
\end{equation}
where the operators are
\begin{eqnarray}
\label{K-simul}
\hat{K}_1^n &=& diag(U_{11}^{nn}, U_{22}^{nn}, U_{33}^{nn}, U_{44}^{nn})+
U_{23}^{nn}(|eg\rangle\langle ge|+h.c.)
\nonumber \\
\hat{K}_2^n &=& \sqrt{2}U_{12}^{nn-1}|ee\rangle\langle
s|+\sqrt{2}U_{24}^{nn-1}|s\rangle\langle gg| 
\nonumber \\
\hat{K}_3^n &=& \sqrt{2}U_{21}^{nn+1}|s\rangle\langle
ee|+\sqrt{2}U_{42}^{nn+1}|gg\rangle\langle s| 
\nonumber \\
\hat{K}_4^n &=& U_{14}^{nn-2}|ee\rangle\langle gg|
\nonumber \\
\hat{K}_5^n &=& U_{41}^{nn+2}|gg\rangle\langle ee|
\end{eqnarray}
with $h.c.$ denoting the Hermitian conjugate and
$|s\rangle=(|eg\rangle+|ge\rangle)/\sqrt{2}$.  Note that $\hat{K}_1^n$
determines the process that does not change the mean energy of the
field while $\hat{K}_2^n$ and $\hat{K}_4^n$ determine, respectively,
one and two photon absorption processes.  $\hat{K}_3^n$ and
$\hat{K}_5^n$, respectively, describe atomic transitions through one and two
photon emission processes.

Each Kraus operator projects the atomic state into a pure state
if the atomic state is initially pure.  If the atoms are
initially in their excited states the only operation that projects the
system into an entangled state is $\hat{K}_3^n$.  For a two qubit system described by the density
operator $\hat{\rho}$, a measure of entanglement can be defined in
terms of the negative eigenvalues of the partial transposition
\cite{LeeKim00}:
\begin{equation}
{\cal E}=-2\sum_i\mu_i^-.
\label{definition-entanglement}
\end{equation}
where $\mu_i^-$ are the negative eigenvalues of the partial
transposition of $\hat{\rho}$.  When ${\cal E}=0$ the two qubits are
separable \cite{Peres,Horodecki} and ${\cal
  E}=1$ indicates maximum entanglement between them.  By substituting
the initial condition of the atoms, {\em i.e.}
$\hat{\rho}_s=\hat{\rho}_s(0)=|ee\rangle\langle ee|$, into
Eq.(\ref{Kraus-representation-1}), we find
\begin{equation}
\label{rho-time}
\hat{\rho}_s(t)=A_e|ee\rangle\langle ee|
+2A_s|s\rangle\langle s|+A_g|gg\rangle\langle gg|
\end{equation}
where $A_e=\sum_n P_n(U_{11}^{nn})^2$, $A_s=\sum P_n|U_{21}^{nn+1}|^2$
and $A_g=\sum_n P_n(U_{41}^{nn+2})^2$.  The eigenvalues of its
partial transposition are then $\mu_o=A_s$ and 
$2\mu_\pm=(A_e+A_g)\pm[(A_e+A_g)^2-4A_eA_g+4A_s^2]^{1/2}$.  It is
obvious that $\mu_o$ and $\mu_+$ are always positive.  The eigenvalue
$\mu_-$ becomes negative if and only if $A_s >
\sqrt{A_eA_g}$.  In order to analyze this case, let us first consider the
possibility of entangling two atoms when the field is in a Fock state
$|\ell\rangle$, {\em i.e.}  $P_n=\delta_{n,\ell}$.  Using the values
of $U_{ij}^{nm}$ in (\ref{unitary-operator}), it is straightforward to
prove that $|U_{21}^{\ell\ell+1}|^2 \leq |U_{11}^{\ell\ell}
U_{41}^{\ell\ell+2}|$ regardless of $\ell$ and interaction time.  It
is interesting to note that when two atoms are initially in their
excited states, they cannot be entangled by simultaneous interaction
with a Fock state whose energy is definite.  In fact, the atoms are
not entangled even via the vacuum ($\bar n=0$) in this case.  For
their interaction with a thermal field, $\sqrt{A_eA_g}$ has a lower
bound $\sum_n P_n|U_{11}^{nn}U_{22}^{mm}|$ which, extending the
analysis for the Fock state, is found to be larger than the $A_s$.  This
proves that when Fock states can not entangle two atoms by
simultaneous interaction, neither can their classical mixture, for example a
thermal state, can do it.

It has been shown that a single two-level atom and a thermal field can be
entangled due to their linear interaction, regardless of the
temperature of the field \cite{Bose01}.  If there are two atoms
interacting with the thermal field, can we still see the entanglement
between the field and any single atom?  As discussed earlier, by interacting with
the thermal field, a pure quantum system becomes mixed, so we know that
the mutual information between the system and thermal field should become
non-zero.  However, this does not mean that there should be atom-field
entanglement because the mutual information may include classical
correlation as well.  In order to consider the entanglement between
a single atom and the field, we first find the single atom-field density operator
$\hat{\rho}_{a-f}(t)$ by tracing the total density operator
$\hat{\rho}(t)$ over the variables of the other atom.  The atom is described as a
two-dimensional system while the field is in infinite-dimensional
space; the density operator $\hat{\rho}_{a-f}(t)$ is thus defined in
a $2\times\infty$ dimensional space.  An analytical form of entanglement
is not available for the system in $2\times\infty$ space, so that Bose
{\it et al} project the atom-field state onto a $2\times 2$ subspace
and then compute the entanglement of formation for each of the
outcomes.  As this projection can be done by local actions, it does
not increase the entanglement and the result gives a lower bound on
the entanglement of the entire atom-field system.  We have calculated the
average of entanglements in this way for projected states, and this is plotted in
Fig. 1 for various values of the mean photon number $\bar n$.  
We see that the atom and field are
always entangled for $t>0$ despite of the presence of the other atom.  After a
little algebra we can analytically prove this as a lower bound of
entanglement occurs in the subspace of $|e,1\rangle, |e,0\rangle,
|g,1\rangle$ and $|g, 0\rangle$ when $P_0^2
|U_{21}^{01}|^2(U_{11}^{00})^2 >0$.  The atom-field entanglement is a
key to the decoherence process.

More interestingly, we find in Fig. 2 that atoms are entangled by a
thermal field when one of the atoms is prepared in its excited state and
the other in the ground state.  If the initial state is $|eg\rangle$,
the atomic system is represented by the density operator:
\begin{eqnarray}
\label{rho-eg}
\hat{\rho}_s(t)&=&\sum_n P_n
\{(U_{22}^{nn}|eg\rangle+U_{23}^{nn}|ge\rangle)(\langle eg|U_{22}^{nn} 
+\langle ge|U_{23}^{nn})
\nonumber \\
&+&|U_{12}^{nn-1}|^2|ee\rangle\langle
ee|+|U_{42}^{nn+1}|^2|gg\rangle\langle gg|\}. 
\end{eqnarray}
One of the partial transposition eigenvalues may be negative
when
$4\sum_{nm}P_nP_m(-|U_{12}^{nn-1}|^2|U_{42}^{mm+1}|^2+U_{23}^{nn}U_{22}^{nn}U_{23}^{mm}U_{22}^{mm})>0$.
Substituting $U_{ij}^{nm}$ in (\ref{unitary-operator}), the left hand
side of the inequality is found to be given by
$(1/4)\sum_{nm}P_nP_m\sin^2\Omega_nt\sin^2\Omega_mt/[(2n+1)(2m+1)]$ 
which is always positive.  It is
surprising that a thermal state, which is a highly chaotic state in an
infinite dimensional Hilbert space, can entangle two qubits depending
on their atomic initial preparation.

Let us now consider the situation when the atoms are initially both in their ground
states.  In this case, $\hat{K}_2^n$ in Eq.(\ref{K-simul}) is the only
operator that projects the atoms into an entangled state.  The density
operator for the atoms has the same form as (\ref{rho-time}) but with
different parameters: $A_e=\sum_n P_n(U_{14}^{nn-2})^2$, $A_s=\sum
P_n|U_{24}^{nn-1}|^2$ and $A_g=\sum_n P_n(U_{44}^{nn})^2$.  The
measure of entanglement (\ref{definition-entanglement}) is calculated
and plotted in Fig. 3, which clearly shows entanglement between the
two atoms for some interaction times.  As we did for the initial state
$|ee\rangle$, we again first consider the case of Fock-state interaction, 
in order to provide insights for the analysis of the more complicated case of thermal-field interaction. For
the interaction of atoms with the Fock state $|\ell\rangle$, one of
the eigenvalues for the partial transposition of $\hat{\rho}_s(t)$ is
negative when $E_g^\ell\equiv|U_{24}^{\ell\ell-1}|^2-
|U_{44}^{\ell\ell} U_{14}^{\ell\ell-2}|> 0$.  Substituting the values
$U_{ij}^{nm}$ in Eq.(\ref{unitary-operator}) into the definition of
$E_g^\ell$, we find
\begin{equation}
\label{criterion}
E_g^\ell=\frac{1-c}{2\ell-1}\left[\ell(1+c)-{1 \over d}|\ell (1+c)-1|\right]
\end{equation}
where $c=\cos\Omega_{n-1}t$ and $d=(2\ell-1)/\sqrt{\ell(\ell-1)}$.
The atoms are not entangled by the vacuum interaction as $E_g^\ell=0$
for $\ell=0$.  We find that the two atoms {\em are} entangled when $c >
c_s\equiv -1+1/(\ell+d)$ for any Fock state ($\ell\geq 1$).  This
means that it is possible to entangle two atoms by the linear
interaction with any Fock state of the cavity field if the atoms are initially prepared in
their ground states, in sharp contrast to the case of the
initial preparation of atoms in the excited states.  In particular,
for the Fock state of $|1\rangle$, the atoms are always entangled
except when 
$\cos\Omega_{\ell-1}t=1$.  The critical parameter $c_s$ takes the
minimum value -1 at $\ell=1$ and maximizes to around $-0.76$ at
$\ell=2$ then gradually decreases as $\ell$ gets larger.  The maximum
value of $E_g^\ell$ is $1/\ell$ at $c=-1+1/\ell$ so that entanglement
is smaller when $\ell$ gets larger.

We now consider the two atoms interacting with a thermal field.  When
the atoms are initially in $|eg\rangle$, entanglement is induced by a
thermal field regardless of its mean photon number.  Is it still true
for the atoms initially in $|gg\rangle$?  We find the answer is no.  
The thermal state is a weighted mixture of Fock states with the weight
$P_\ell$ given by Eq.(\ref{thermal-density}). The weight factor
$P_\ell$ is a decreasing function with $\ell$.  When $\bar n$ is very
small, because of the vacuum dominance the atoms are not very much
entangled.  As $\bar n$ gets moderately larger, $P_1$ becomes more
pronounced.  Because $|1\rangle\langle 1|$ gives strong atomic
entanglement, the atoms are entangled more in this case.  However, as
$\bar n\gg 1$, the weight function becomes flat and cancelation
between the component Fock states washes out the
entanglement feature.

So far we have considered the case when the atoms are in their pure
states.  Bose {\em et al.} \cite{Bose01} comment that the purity of
the qubit is an important ingredient to see entanglement between the
qubit and a massive system but did not show the dynamics of the
entanglement when the qubit is initially prepared in its mixed state.
It is not possible to analyze the atom-field entanglement when the
atoms are in a mixed state because of the lack of tools to analyze
such a mixed $2\times\infty$ system.  However, we can analyze how the
initial mixedness of atoms affects the entanglement between the atoms.
We assume that each atom is initially prepared in a thermal mixture
so that their initial density operators are
\begin{equation}
\label{mixed-atom}
\hat{\rho}_s=\Pi_{i=1,2}[\lambda|e\rangle_i\langle e|+(1-\lambda)|g\rangle_i\langle g|]
\end{equation}
where $\lambda$ depends on the temperature of the atoms.  Using the
Kraus representation (\ref{Kraus-representation-1}) for the initial
condition (\ref{mixed-atom}), the evolution of the density operator for
the two atoms is found in the form:
$\hat{\rho}_s(t)=A_e|ee\rangle\langle ee|+ A_s|s\rangle\langle
s|+A_a|a\rangle\langle a|+A_g|gg\rangle\langle gg|$ where
$|a\rangle=(|eg\rangle+|ge\rangle)/\sqrt{2}$ and $A_{e,s,a,g}$ are
time-dependent coefficients.  The atoms are then entangled when
$(A_s-A_a)-2\sqrt{A_eA_g}>0$.  The measure of entanglement is plotted
for the interaction of atoms with a low temperature field with $\bar
n=1$.  It is remarkable to see that with even a very small amount of
mixture with $\lambda=0.065$, atomic entanglement is nearly washed
out.

In conclusion, we have demonstrated the very interesting result that two atoms can become
entangled through their interaction with a highly chaotic system depending on the initial
preparation of the atoms.  This study provides a degree of analytical
understanding of the decoherence mechanism for a quantum system composed
of a few qubits when the reservoir with which they interact retains some memory \cite{Palma}.


This work was supported by the UK Engineering and Physical Science
Research Council
and by the Korean Ministry of Science and Technology
through the Creative Research Initiatives Program under Contract No.
00-C-CT-01-C-35.

\begin{figure}
  \begin{center}
    \includegraphics[width=0.4\textwidth]{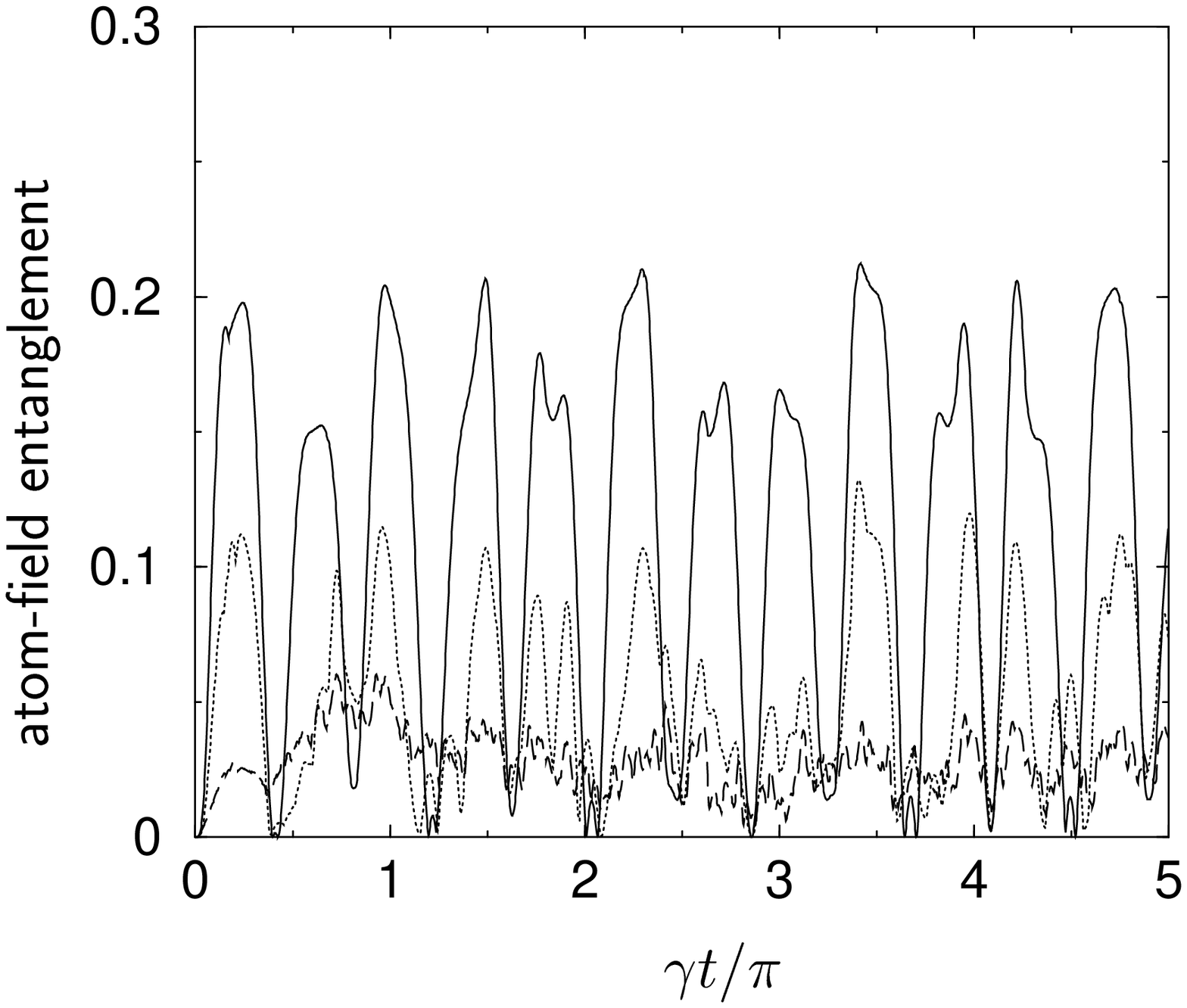}
    \caption{Single atom-field entanglement against the interaction time
      when the pair of atoms is initially prepared in the state $|ee\rangle$
      for $\bar n=0.1$ (solid), 1.0 (dotted) and 10.0 (dashed). The
      entanglement always exits for all times $t>0$.  }
     \end{center}
\end{figure}

\begin{figure}
  \begin{center}
    \includegraphics[width=0.4\textwidth]{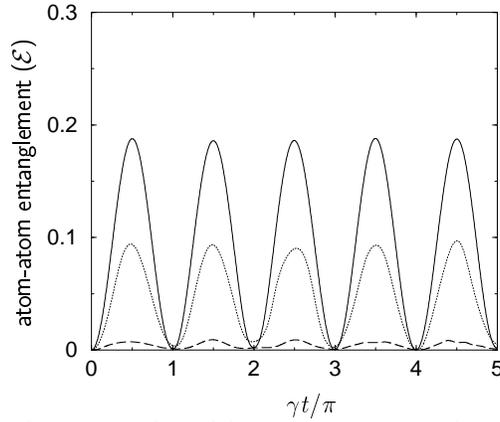}
    \caption{Atom-atom entanglement induced by interaction with a thermal
      field when the atoms are initially prepared in $|eg\rangle$ for $\bar
      n=0.1$ (solid), 1.0 (dotted) and 10.0 (dashed).  }
     \end{center}
\end{figure}

\begin{figure}
  \begin{center}
    \includegraphics[width=0.4\textwidth]{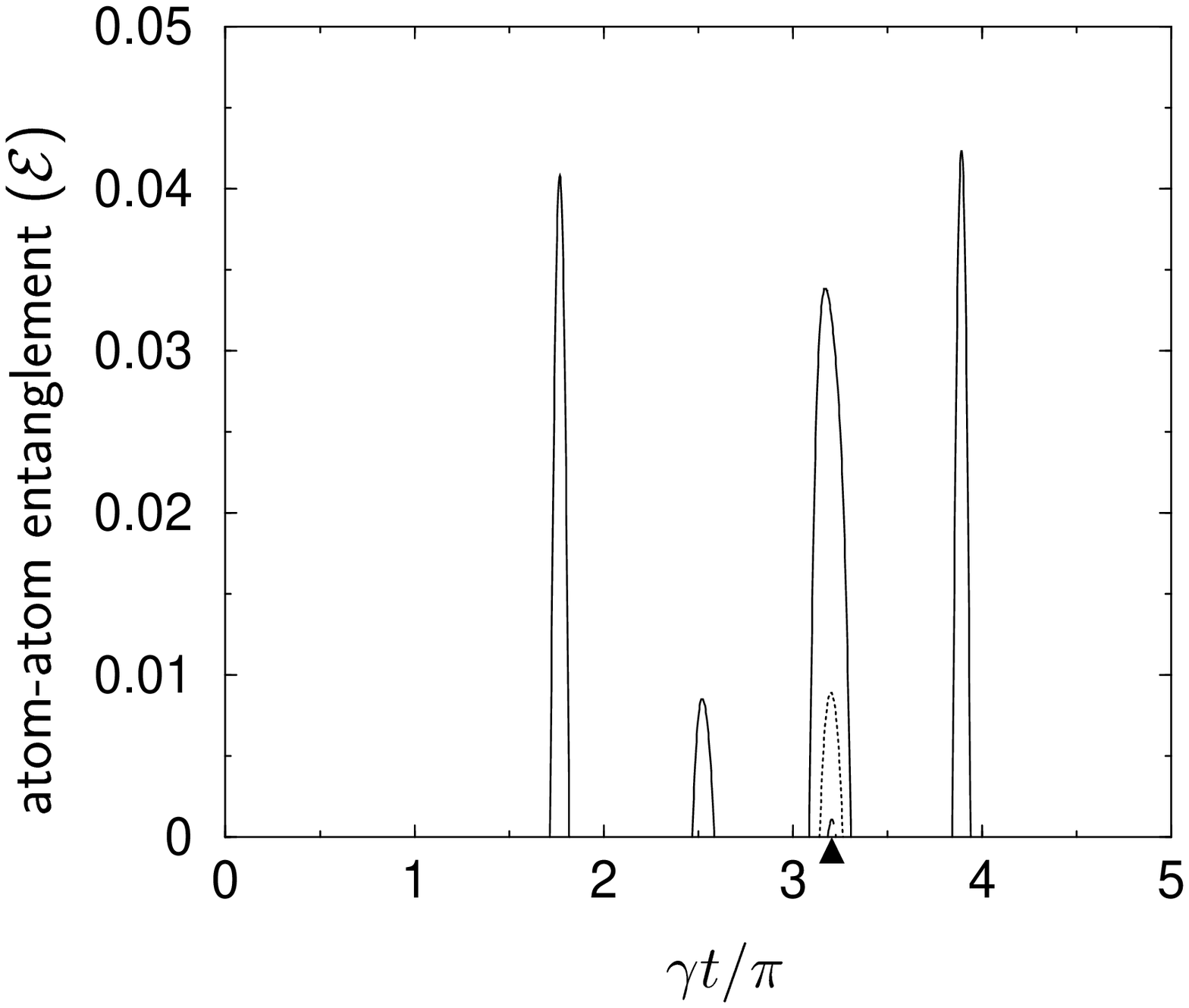}
    \caption{Atom-atom entanglement induced by interaction with a thermal
      field of $\bar n=1$ when the atoms are initially in a pure state
      ground state (solid), {\em i.e.} $\lambda=0$ in
      Eq.(\ref{mixed-atom}), and in a mixed state of $\lambda =0.05$
      (dotted). In the case of $\lambda=0.065$, entanglement is shown
      by a dashed line at the spot indicated by a triangle.  }
  \end{center}
\end{figure}

\end{document}